\documentstyle[fleqn]{tp}

\input epsf

\def\eqref#1{(\ref{eqn:#1})}

\def\hmpc{h^{-1}{\rm Mpc}}
\def\kms{{\rm \;km\;s^{-1}}}
\def\invkms{({\rm km\;s}^{-1})^{-1}}

\def\nh1{n_{\rm HI}}
\def\rhobar{{\overline\rho}}
\def\overden{{\rho/\rhobar}}
\def\lya{Ly$\alpha$\ }
\def\la{\mathrel{\hbox{\rlap{\hbox{\lower4pt\hbox{$\sim$}}}\hbox{$<$}}}}
\def\ga{\mathrel{\hbox{\rlap{\hbox{\lower4pt\hbox{$\sim$}}}\hbox{$>$}}}}
\def\K{{\rm K}}
\def\apj{ApJ}
\def\aj{AJ}

\def\mnras{MNRAS}
\def\ol{\Omega_\Lambda}
\def\om{\Omega_0}
\def\omb{\Omega_{\rm b}}

\begin{document}

\twocolumn[
\title{Cosmology with the Lyman-alpha Forest}
\author{David H. Weinberg$^1$, 
Scott Burles$^2$, 
Rupert A. C. Croft$^{1,3}$,
Romeel Dav\'e$^{4,5}$,\\
Gilberto Gomez$^{6,7}$,
Lars Hernquist$^{3,4}$, 
Neal Katz$^8$, 
David Kirkman$^9$,\\
Shulan Liu$^8$,
Jordi Miralda-Escud\'e$^{10}$,
Max Pettini$^{11}$,\\ 
John Phillips$^1$, 
David Tytler$^9$, 
Jason Wright$^{12}$\\
{\it $^1$Ohio State University, Columbus, Ohio, USA}\\
{\it $^2$University of Chicago, Chicago, Illinois, USA}\\
{\it $^3$Harvard-Smithsonian CfA, Cambridge, Massachusetts, USA}\\
{\it $^4$U.C. Santa Cruz, Santa Cruz, California, USA}\\
{\it $^5$Princeton University Observatory, Princeton, New Jersey, USA}\\
{\it $^6$University of Wisconsin, Madison, Wisconsin, USA}\\
{\it $^7$Instituto de Astronom\'ia, UNAM, Mexico}\\
{\it $^8$University of Massachusetts, Amherst, Massachusetts, USA}\\
{\it $^9$U.C. San Diego, San Diego, California, USA}\\
{\it $^{10}$University of Pennsylvania, Philadelphia, Pennsylvania, USA}\\
{\it $^{11}$Royal Greenwich Observatory, Cambridge, UK}\\
{\it $^{12}$Boston University, Boston, Massachusetts, USA}
}
\vspace*{16pt}   

ABSTRACT.\
We outline the physical picture of the high-redshift \lya forest that has 
emerged from cosmological simulations, discuss statistical characteristics of 
the forest that can be used to test theories of structure formation, present a 
preliminary comparison between simulation predictions and measurements from a
sample of 28 Keck HIRES QSO spectra, and summarize the results of a recent
determination of the linear mass power spectrum $P(k)$ at $z=2.5$ from a sample
of 19 moderate resolution QSO spectra.  The physical picture is 
simple if each QSO spectrum is viewed as a continuous non-linear map of the
line-of-sight density field rather than a collection of discrete absorption 
lines.  To a good approximation, the relation between \lya optical depth and
mass overdensity is $\tau=A(\overden)^{1.6}$.  The constant $A$ depends on 
poorly known physical parameters, but for a specified cosmological model its
value can be calibrated using the mean flux decrement $\langle D \rangle \equiv
\langle 1-e^{-\tau} \rangle$, leaving all other statistical properties of the
forest as independent model predictions.  The distribution of flux decrements
is closely tied to the probability distribution function (PDF) of the
underlying mass distribution, and hence to the amplitude and PDF (Gaussian or
non-Gaussian) of the primordial density fluctuations.  The threshold crossing
frequency, analogous to the 3-d genus curve,
responds to the shape and amplitude of the
primordial $P(k)$ and to the values of $\om$ and $\ol$.  The predictions of
open CDM and $\Lambda$CDM models agree well with the measured
flux decrement distribution at smoothing lengths of $25\kms$ and $100\kms$
and with the threshold crossing frequency at $100\kms$.  Discrepancy with the 
observed threshold crossing frequency at $25\kms$ smoothing 
may reflect the combined effects of noise in the data and
limited mass resolution of the simulations.  
The moderate resolution spectra yield the slope and amplitude of the linear
mass $P(k)$ at $2\pi/k \sim 700\kms$.  The slope, never previously
measured on these scales, agrees with the predictions of inflation+CDM models.
Combining the amplitude with COBE normalization imposes a constraint
on these models of the form $\om h^\alpha n^\beta \omb^\gamma = {\rm constant}.$
Assuming Gaussian primordial fluctuations and
a power spectrum shape parameter $\Gamma \approx 0.2$, consistency of the 
measured $P(k)$ with the observed cluster mass function at $z=0$ requires
$\om=0.46^{+0.12}_{-0.10}$ for $\ol=0$ and $\om=0.34^{+0.13}_{-0.09}$ for
$\ol=1-\om$ ($1\sigma$ errors).
\endabstract]

\markboth{David H. Weinberg et al.}{Cosmology with the \lya Forest}

\small

\section{Introduction}
For decades, the study of large scale structure was virtually synonymous
with the study of galaxy clustering.  The 1980s and 1990s have seen
spectacular improvements in galaxy redshift 
and peculiar velocity data, many of them documented at this meeting.
At this point, our ability to draw cosmological conclusions from these
data is often limited less by their statistical uncertainties than by
our limited understanding of the complex physics of galaxy formation.
CMB anisotropy measurements have been enormously influential because
their underlying physics is simpler, allowing a straightforward connection
between theory and observation.
However, even the ambitious CMB experiments now underway will yield
only a particular projection of the evolution of cosmic structure,
mostly concentrated at $z \sim 1000$.
Other measures that probe 3-dimensional structure at later redshifts
are needed to complement these experiments.

The success of hydrodynamic cosmological simulations in explaining
the basic properties of the \lya forest
(Cen et al.\ 1994; Zhang, Anninos, \& Norman 1995; Hernquist et al.\ 1996;
Wadsley \& Bond 1996; Theuns et al.\ 1998) has opened up a new
testing ground for cosmological theories.
Ground-based observations yield superb data on the \lya forest at
$z\sim 2-4$, and HST spectra provide lower resolution measurements
at $z \sim 0-2$.
Equally important, the simulations lead to a simple
physical picture of the \lya forest, one that is well approximated
by semi-analytic models that were to a significant
extent developed prior to and independent of the simulations
(McGill 1990; Bi 1993; Bi, Ge, \& Fang 1995; Bi \& Davidsen 1997;
Hui, Gnedin, \& Zhang 1997).
We will briefly describe this physical picture in the next section
(more extended discussions along these lines appear in
Weinberg et al.\ 1997 and Weinberg, Katz, \& Hernquist 1998,
hereafter WKH).  We will then discuss the physical significance
of two statistics that can be used to test cosmological models
against \lya forest data and present a preliminary comparison
between predictions from smoothed particle hydrodynamics (SPH)
simulations and measurements from a sample of Keck HIRES spectra.
We conclude with a discussion of
a recent determination of the mass power spectrum at $z=2.5$
and its implications for the value of $\om$ and the parameters
of CDM models.

\section{Physics of the \lya Forest}
\begin{figure*}
\centerline{
\epsfxsize=3.6truein
\epsfbox[60 470 555 735]{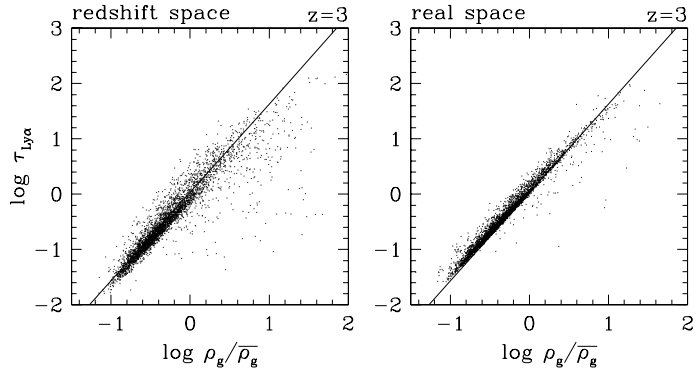}
}
\caption[]{
\label{fig:fgpa}
A test of the Fluctuating Gunn-Peterson Approximation
on artificial spectra extracted from an SPH simulation of
standard CDM in real space (right) and redshift space (left),
at $z=3$.  Points show the relation between \lya optical depth and gas 
overdensity in the simulated spectra, pixel by pixel.  
Diagonal lines show the prediction of equation~\eqref{fgpa}.
}
\end{figure*}

In hydrodynamic simulations of high redshift structure ($z \sim 2-5$),
the \lya forest is produced mainly by gas with overdensity $\overden \la 10$.
Most of this gas is unshocked, and the competition between photoionization
heating and adiabatic cooling leads to a tight relation 
between $\rho$ and $T$ that is well approximated by a power law,
$T = T_0(\overden)^\alpha$.  The values of $T_{0}$ and $\alpha$ depend on 
the reionization history of the universe and on the spectral shape of 
the UV background; they typically lie in the ranges
$4000\,\K \la T_0 \la 10,000\,\K$ and  $0.3 \la \alpha \la 0.6$
(Hui \& Gnedin 1997).  
The optical depth for \lya absorption is proportional to the
neutral hydrogen density, which for gas in photoionization equilibrium
near $10^4\K$ is proportional to $\rho^2 T^{-0.7}/\Gamma_{\rm HI}$,
where
$\Gamma_{\rm HI}$ is the HI photoionization rate and the $T^{-0.7}$
factor accounts for the temperature dependence of the recombination rate.
The combination of photoionization equilibrium and the $\rho-T$ relation
therefore leads to a power law relation between optical
depth, $\tau$, and gas overdensity $\overden$, 
\begin{eqnarray}
\tau & = & A(\overden)^\beta\; , \label{eqn:fgpa} \\
A & = & 0.835
        \left(\frac{1+z}{4}\right)^6
        \left(\frac{\omb h^2}{0.02}\right)^2 \;\times \nonumber \\
      &&\left(\frac{h}{0.65}\right)^{-1} 
        \left(\frac{H(z)/H_0}{4.46}\right)^{-1} \;\times \nonumber \\
      &&\left(\frac{\Gamma_{\rm HI}}
		   {10^{-12}\;{\rm s}^{-1}}\right)^{-1} 
        \left(\frac{T_0}{10^4\;{\rm K}}\right)^{-0.7} 
      \nonumber
\end{eqnarray}
where $\beta \equiv 2-0.7\alpha \approx 1.6$.

Since equation~\eqref{fgpa} describes the analog of Gunn-Peterson (1965)
absorption for a non-uniform, photoionized medium, we refer to it
as the Fluctuating Gunn-Peterson Approximation, or FGPA
(see Rauch et al.\ 1997; Croft et al.\ 1998a; WKH).
The FGPA assumes that all gas lies on the temperature-density relation,
and in the form expressed here it ignores the effects of peculiar
velocities and thermal broadening.
In principle, the quantity $\overden$ in equation~\eqref{fgpa}
represents the {\it gas} overdensity.  However, in the cool, low
density regions that are responsible for most \lya forest absorption,
pressure gradients are usually small compared to gravitational forces.
The gas and dark matter therefore trace each other quite well,
with the gas distribution being slightly smoother in the neighborhood
of low overdensity peaks of the dark matter distribution
(Gnedin \& Hui 1998; Bryan et al.\ 1998).

Figure~\ref{fig:fgpa} tests the FGPA against results from an SPH
simulation of ``standard'' CDM
(hereafter SCDM, with $\Omega=1$, $h=0.5$, $\sigma_8=0.7$), at $z=3$.
The right hand panel plots the \lya optical depth against gas
overdensity for spectra extracted in real space, i.e., with thermal
broadening and peculiar velocities set to zero.  Most points lie
on a tight, clearly defined sequence.  Points lying well below this
sequence come from regions where shock heating has raised the gas 
temperature above the power law $\rho-T$ relation, depressing
the recombination rate and hence the neutral hydrogen optical depth.
Peculiar velocities and thermal broadening increase the scatter
of the relation in redshift space (left hand panel), but they do not 
destroy it.  The diagonal lines show the prediction of 
equation~\eqref{fgpa}, with all ``free'' parameters
($\omb$, $h$, $T_0$, $\Gamma_{\rm HI}$, $\beta$, $H(z)/H_0$)
set to the values that they have in the simulation.
The FGPA provides a good but not perfect description of the numerical
results, breaking down mainly in the regions where shock heating
is important.

For an approximate numerical approach that is much
cheaper than a high resolution hydrodynamic calculation, one can
run a lower resolution particle-mesh (PM) N-body simulation, 
compute the density field from the evolved particle distribution, 
impose the $\rho-T$ relation, and extract spectra.  
This technique uses a fully non-linear
solution for the density and velocity fields, but it still assumes
that gas traces dark matter and that all gas lies on the $\rho-T$  
relation, approximations that are good but not perfect.
As shown in WKH (Figure 8), the agreement between the PM approximation
and a full hydrodynamic simulation is very good over most of the spectrum, 
but it breaks down in some higher density, shocked regions.
Similar techniques have been used by Gnedin \& Hui (1998), who
also incorporate an approximate treatment of gas pressure in the
N-body calculation, and by Petitjean, M\"ucket, \& Kates (1995)
and M\"ucket et al.\ (1996), who use a heuristic method to incorporate
shock heating.

\section{Continuous Field Statistics}

The approximate physical picture just outlined implies a simple and
direct relation between observed flux and mass overdensity,
\begin{equation}
\frac{F}{F_c} = e^{-\tau} = e^{-A(\rho/\rhobar)^{1.6}}. 
\label{eqn:flux}
\end{equation}
The unabsorbed continuum flux, $F_c$, must be estimated by
fitting a continuum to regions of low apparent absorption or by
extrapolating the continuum from the red side of the \lya emission line.
The constant $A$ depends on poorly known parameters,
but given a cosmological model that predicts the statistical properties
of the density and peculiar velocity fields, the value of $A$ can
be fixed by matching one observable, the most obvious choice being the
mean flux decrement $\langle D \rangle \equiv \langle 1-e^{-\tau}\rangle$.
This mean decrement normalization calibrates the one
unknown parameter that determines
the relation between mass fluctuations and optical depth fluctuations.

The traditional approach to analysis of the \lya forest is to decompose 
each spectrum into lines and compute the statistical distributions of 
the line properties.  However, while the relation between flux and mass density 
is simple, the relation between the number of lines and mass density is not.
Line decomposition also interposes a complicated, non-linear algorithm 
in the path between data and model predictions, an algorithm whose
performance depends in an intricate way on the
signal-to-noise ratio and spectral resolution of the data.
Statistical measures that treat the spectrum as a continuous field
are likely to be more powerful than line decomposition methods for
discriminating cosmological models, because they involve minimal manipulation
of the observational data and because they are better attuned
to the underlying physics of the absorbing medium.

\begin{figure*}
\centerline{
\epsfxsize=3.6truein
\epsfbox[60 220 555 735]{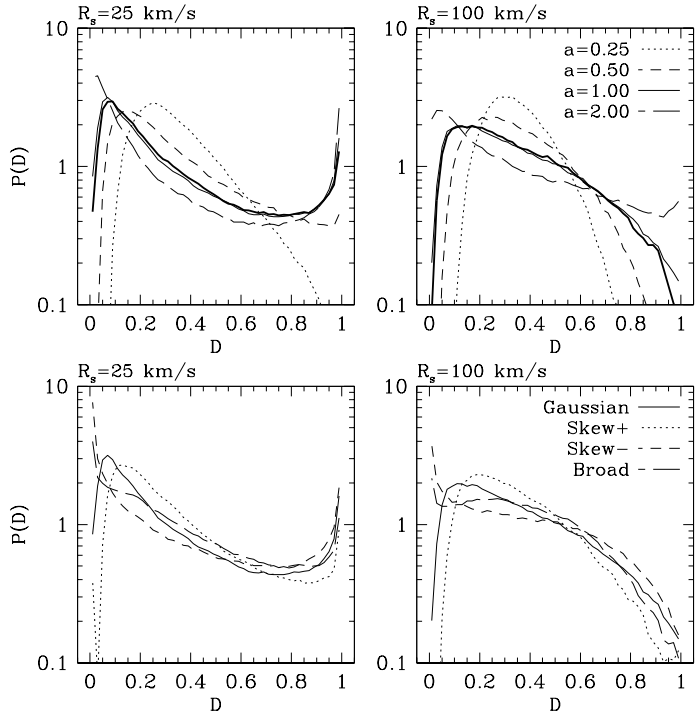}
}
\caption[]{
\label{fig:fdfphys}
The influence of mass fluctuation amplitude and primordial PDF shape
on the flux decrement distribution function, $P(D)$, with smoothing lengths
of $25\kms$ (left) and $100\kms$ (right).  In all panels, light
solid lines show results from a PM simulation of the SCDM model
with Gaussian initial conditions.
In the upper panels, heavy solid lines show the results from a full
SPH simulation with the same initial conditions, and other line
types show results from PM simulations in which the amplitude of the
initial conditions has been scaled by the factor {\rm a} 
indicated in the legend.
Lower panels compare SCDM to models with the same
initial power spectrum but
non-Gaussian PDFs that are skew-positive, skew-negative,
or symmetric with broad tails.
}
\end{figure*}

\begin{figure*}
\centerline{
\epsfxsize=3.6truein
\epsfbox[60 220 555 735]{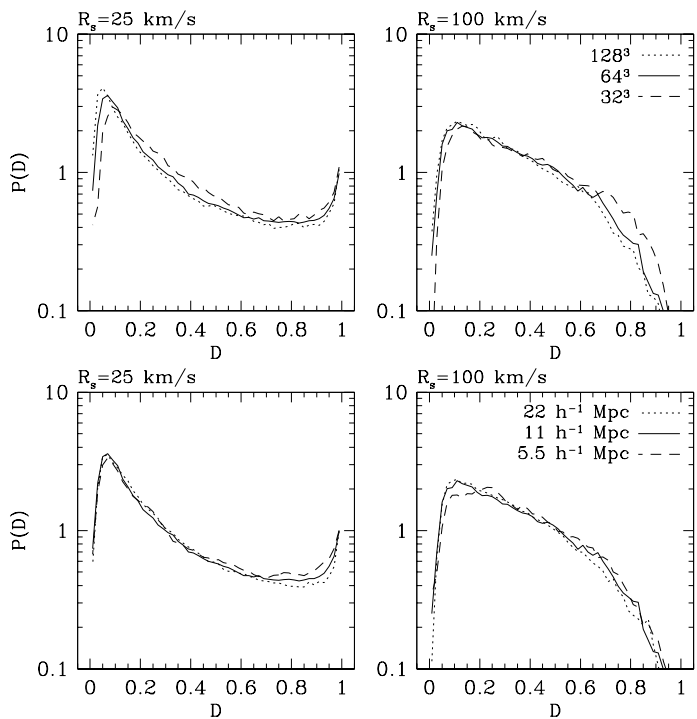}
}
\caption[]{
\label{fig:fdfres}
The influence of mass resolution and simulation box size on calculations 
of the flux decrement distribution function, in the PM approximation at $z=3$.
Solid lines show results from an LCDM simulation (see Table~\ref{tab:1}
for cosmological parameters) with $64^3$
particles in an $11\hmpc$ comoving box.
In the upper panels,
dotted and dashed lines show results from simulations with the
same box size but $128^3$ and $32^3$ particles, respectively.
In the lower panels, dotted and dashed lines show results from
simulations with the same mass resolution but a
$22\hmpc$ box ($128^3$ particles) and a $5.5\hmpc$ box ($32^3$ particles),
respectively.
}
\end{figure*}

\begin{figure*}
\centerline{
\epsfxsize=3.6truein
\epsfbox[60 470 555 735]{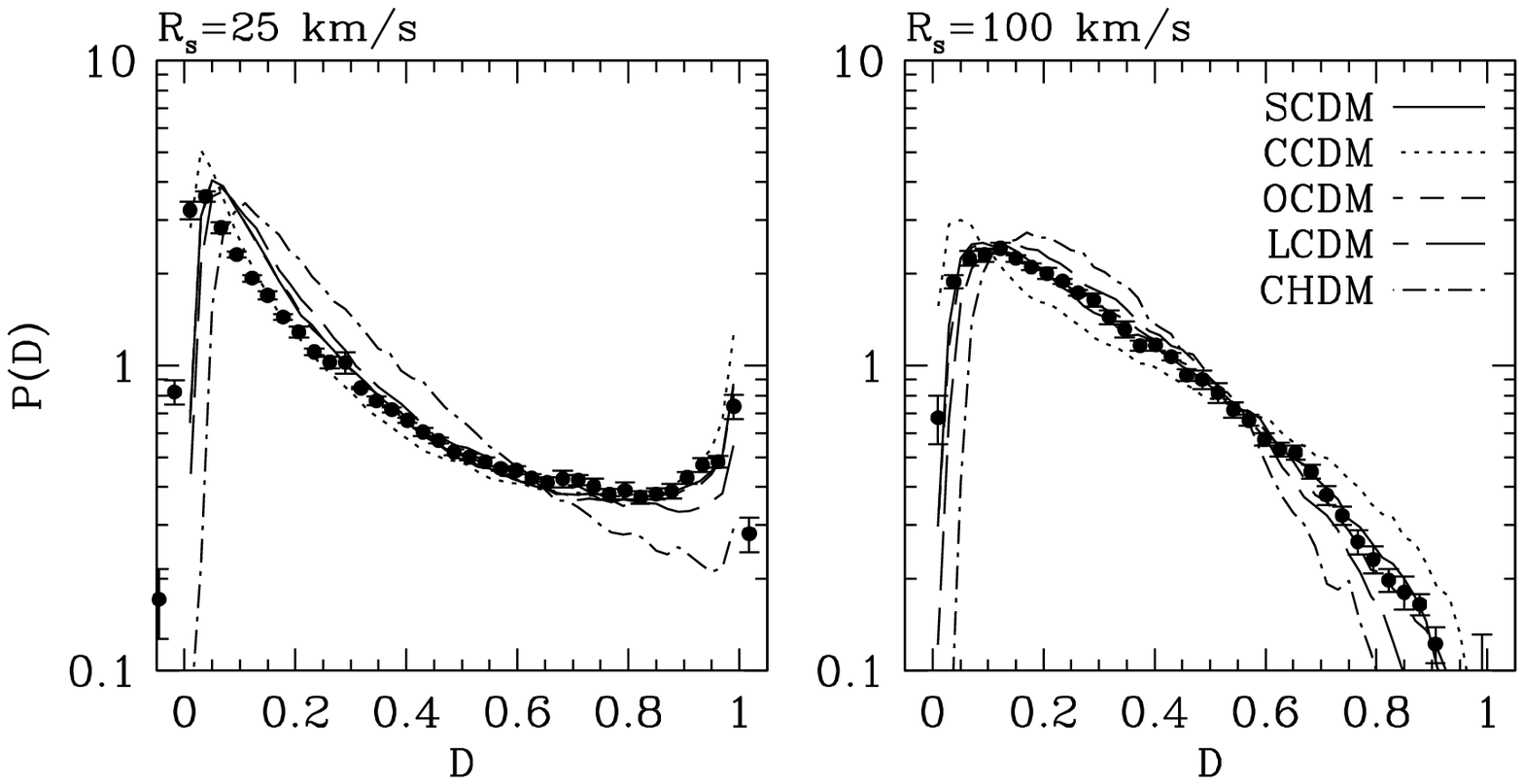}
}
\caption[]{
\label{fig:fdfdata}
Comparison of the flux decrement distribution from SPH simulations
of five different cosmological models to measurements from a
sample of Keck HIRES spectra with mean redshift $z=2.96$.
Parameters of the models are given in Table~\ref{tab:1}.
The resolution and box size of the simulations correspond to the
solid lines in Figure~\ref{fig:fdfres}.  
}
\end{figure*}

One of the simplest of these continuous field statistics is 
the distribution function of
flux decrements $P(D)$, where $D = 1-e^{-\tau}$ and $P(D)dD$ is the
fraction of pixels with flux decrement in the
range $D \longrightarrow D+dD$ 
(Miralda-Escud\'e et al.\ 1996, 1997; Croft et al.\ 1997; Kim et al.\ 1997;
Rauch et al.\ 1997).
This statistic is analogous
to the galaxy counts-in-cells distribution frequently used in studies
of large scale structure.
$P(D)$ can be measured directly from unsmoothed spectra,
but one can also investigate larger scale clustering by smoothing
the spectra and then measuring $P(D)$, just as one would
measure galaxy count distributions in cells of different sizes.
Equation~\eqref{flux} implies that $P(D)$ should be closely related
to the probability distribution function (PDF) of the underlying
mass fluctuations.  For models in which the primordial fluctuations
have a Gaussian PDF, as predicted by most versions of inflation,
the PDF of the non-linear mass fluctuations depends mainly on
the amplitude of the linear power spectrum.
Non-Gaussianity of the primordial fluctuations should also affect
the mass PDF, and hence the flux decrement distribution.

Figure~\ref{fig:fdfphys} confirms these expectations.
In the upper panels, the heavy and light solid lines compare 
the flux decrement distributions measured from a full SPH
simulation of the SCDM model and a PM simulation with the
same initial conditions.  This comparison indicates that the
PM approximation performs quite well for this statistic at these
smoothing lengths (25 and $100\kms$).
The other lines in the upper panels show results from PM simulations
in which the initial mass fluctuations were multiplied by a factor
of 0.25 (dotted), 0.5 (short-dashed), and 2.0 (long-dashed) prior
to evolution.  In all cases the spectra are normalized to the
same mean flux decrement.
As expected, $P(D)$ becomes steadily
broader as the mass fluctuation amplitude increases.
The lower panels show the effect of changing the 
PDF of the primordial fluctuations
while keeping the power spectrum fixed.
We generate non-Gaussian initial conditions for the simulations
using the non-linear transformation method of Weinberg \& Cole (1992).
The skew-negative and broad models, both of which have extended non-Gaussian
tails of negative fluctuations, produce
\lya forest spectra with large numbers of low absorption pixels.
The skew-positive model, by contrast, predicts very few transparent
regions and a flux decrement distribution that is more narrowly
peaked about its mean value.

Using Eulerian hydrodynamic simulations and SPH simulations, respectively,
Bryan et al. (1998) and Theuns et al. (1998) have examined the effects 
of mass resolution and simulation box size on predictions of the
unsmoothed flux decrement distribution.  Both groups find that this
quantity is fairly but not completely robust to changes in numerical 
parameters.  Figure~\ref{fig:fdfres} presents a similar investigation
at smoothing lengths of $25$ and $100\kms$, using the PM approximation.
The speed and simplicity of the PM approximation make it a useful
exploratory tool for such a study, although its convergence properties
will not be identical to those of full hydrodynamic simulations
because it ignores the influence of gas pressure.  
Figure~\ref{fig:fdfres} suggests that the parameters of our current
SPH simulations ($64^3$ particles, $11\hmpc$ box)
are probably sufficient to give accurate predictions
of the flux decrement distribution at these smoothing scales,
although a factor of eight decrease in the resolution or box volume
would change the predictions noticeably.

\begin{table*}
\caption[]{Parameters of the cosmological models shown in 
  Figures~\ref{fig:fgpa}--\ref{fig:dcrossdata}. 
  By definition, $\om = \Omega_{\rm CDM}+\Omega_\nu+\Omega_{\rm b}$.}
  \centering
  \begin{tabular}{l|*{8}{c}}
    \hline
    Model & $\om$ & $\ol$ & $\Omega_\nu$ & $n$ & $h$ & 
	  $\sigma_8$ \\
    \hline\hline
    SCDM          & 1.0 & 0.0 & 0.0 & 1.00 & 0.50 & 0.70 \\
    CCDM          & 1.0 & 0.0 & 0.0 & 1.00 & 0.50 & 1.20 \\
    OCDM          & 0.4 & 0.0 & 0.0 & 1.00 & 0.65 & 0.75 \\
    LCDM          & 0.4 & 0.6 & 0.0 & 0.93 & 0.65 & 0.79 \\
    CHDM          & 1.0 & 0.0 & 0.2 & 1.00 & 0.50 & 0.70 \\
  \end{tabular}
\label{tab:1}
\end{table*}

Figure~\ref{fig:fdfdata} compares
the flux decrement distributions predicted by
SPH simulations of a variety of CDM models 
(with parameters listed in Table~\ref{tab:1})
to measurements from a sample of 28 Keck HIRES spectra.
The absorption redshifts in the observational sample range 
from $z=2.5$ to $z=3.7$, with mean $z=2.96$, and the total path
length in this redshift range
is equivalent to 21 full \lya to Ly$\beta$ regions.
The spectra have been obtained by three of us (SB, DK, DT) for a variety of
purposes, including studies of the primordial deuterium abundance
(e.g., Burles \& Tytler 1998),
the statistical properties of \lya forest lines
(Kirkman \& Tytler 1997a),
and heavy element enrichment in the \lya forest
(Kirkman \& Tytler 1997b).
Error bars on the data points are estimated by dividing the
sample into subsets.
The statistical uncertainties are small because there are many
independent resolution elements in the sample spectra, but the
errors are not independent.
Furthermore, this is a comparison between
results from noiseless simulation data for which the true continuum is
known and results from observed spectra that have noise and a 
locally fitted continuum.  
Substantially more work is required to go from this preliminary
analysis to a definitive comparison between models and data.

Taking the results of Figure~\ref{fig:fdfdata} at face value,
there appears to be fairly good agreement between the observed $P(D)$
and the predictions of the 
SCDM, OCDM, and LCDM models.  These three models all have 
Gaussian initial conditions and a similar mass fluctuation amplitude
on these scales at this redshift, so they
predict similar flux decrement distributions.
There is some discrepancy with the data for low flux decrements
at $25\kms$ smoothing, but this is the regime where noise will have
the largest effect on the measurements.
The COBE-normalized, $\Omega=1$ model, CCDM, seems to have too high
a fluctuation amplitude, predicting a flux decrement distribution
that is too broad at $100\kms$.
The cold+hot dark matter model, CHDM, seems to have fluctuations that are
too weak, predicting an excessively narrow distribution.

\begin{figure*}
\centerline{
\epsfxsize=3.6truein
\epsfbox[60 220 555 735]{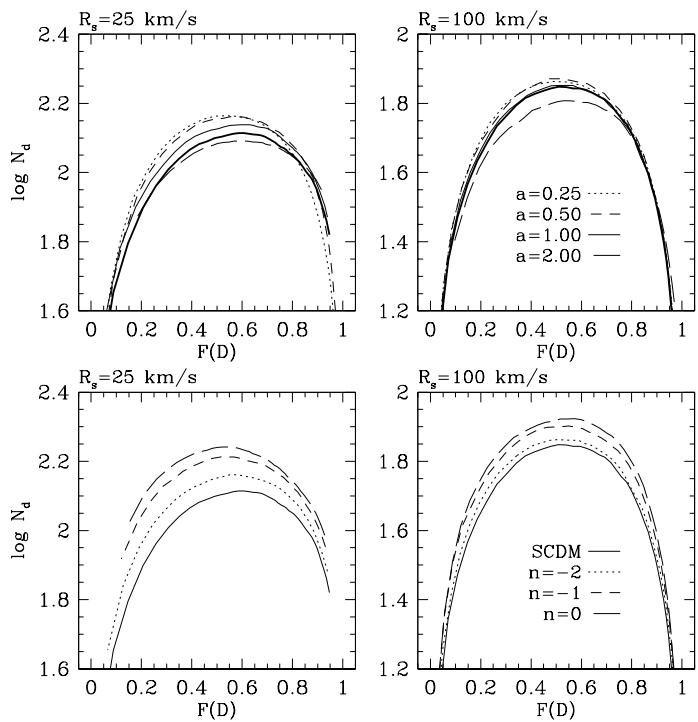}
}
\caption[]{
The influence of the amplitude and shape of the primordial power spectrum
on the threshold crossing frequency.  $N_{\rm d}$, the number of times per
unit redshift that a spectrum crosses a flux decrement threshold $D$ in
the downward direction, is plotted against $F(D)$, the fraction of
pixels with flux decrement less than $D$.  Upper panels show 
models with an SCDM $P(k)$ shape and varying amplitudes;
the heavy line is from an SPH simulation
with ${\rm a}=1.0$ 
and the lighter lines from the PM approximation.  Lower panels
show results from simulations that have the same initial fluctuation
amplitude at a Gaussian smoothing scale of $0.28\hmpc$ (comoving)
but have different $P(k)$ shapes.
\label{fig:dcrossphys}
}
\end{figure*}

\begin{figure*}
\centerline{
\epsfxsize=3.6truein
\epsfbox[60 220 555 735]{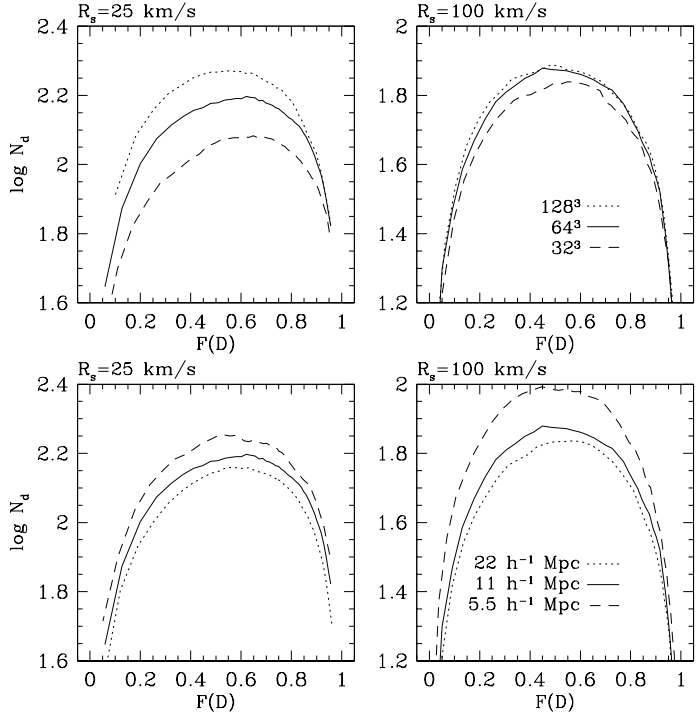}
}
\caption[]{
\label{fig:dcrossres}
The influence of mass resolution and simulation box size on calculations
of the threshold crossing frequency, in the PM approximation.
The quantities plotted are the same as those in 
Figure~\ref{fig:dcrossphys}.  The simulations (LCDM with varying 
particle numbers and box sizes) are the same
as those in Figure~\ref{fig:fdfres}.
}
\end{figure*}

\begin{figure*}
\centerline{
\epsfxsize=3.6truein
\epsfbox[60 470 555 735]{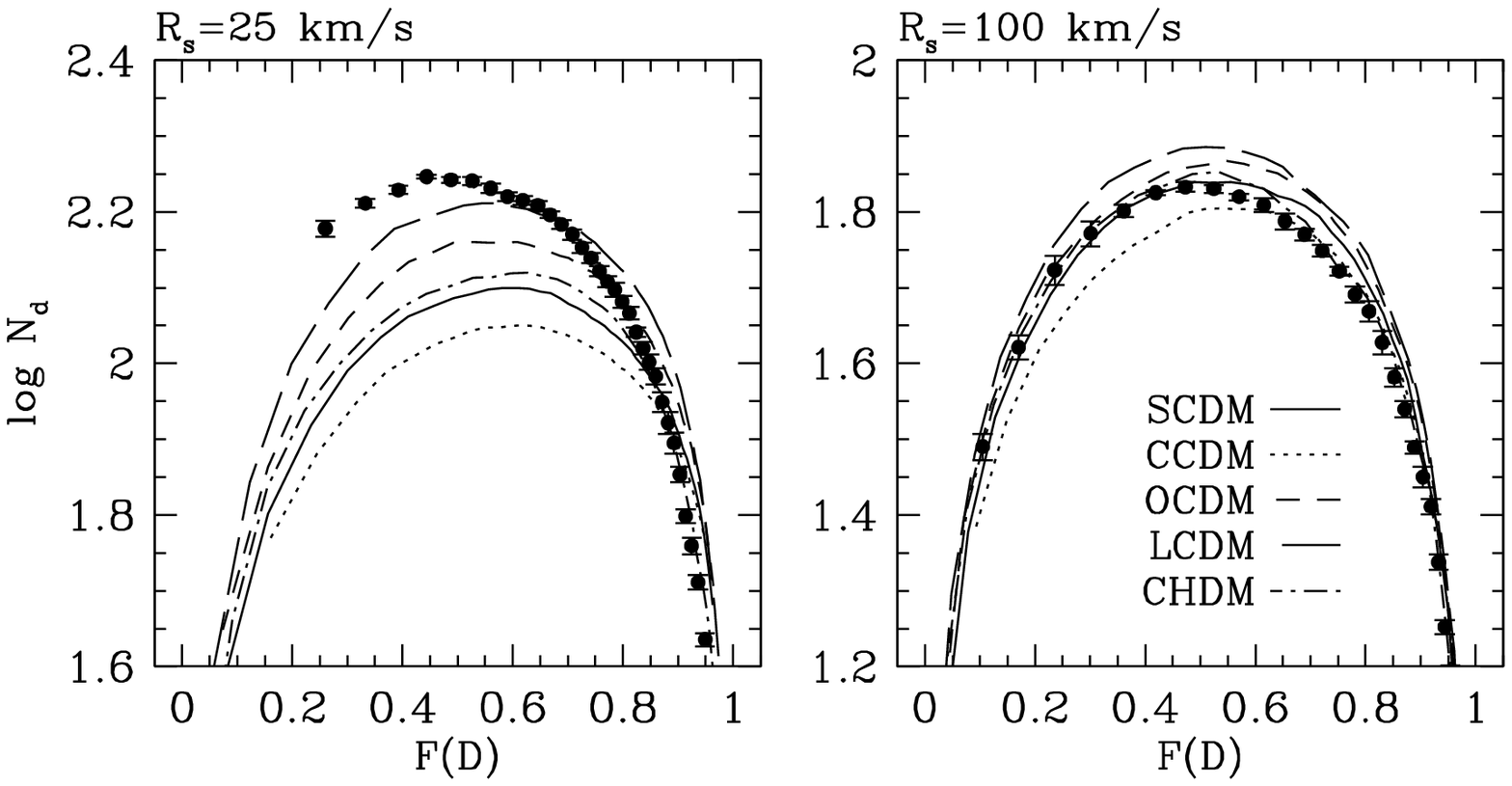}
}
\caption[]{
\label{fig:dcrossdata}
Comparison of the threshold crossing frequency from SPH simulations
of five different cosmological models to measurements from the
Keck HIRES sample.
}
\end{figure*}

Another simple continuous field statistic is the threshold crossing frequency
$N_{\rm d}$, the average number of times per unit redshift that the spectrum
crosses a specified flux decrement threshold
(Miralda-Escud\'e et al.\ 1996; Croft et al.\ 1997; Kim et al.\ 1997).
This statistic is analogous to the ``genus curve'' used to characterize
the topology of the galaxy density field.
Figure~\ref{fig:dcrossphys} uses the PM approximation to illustrate
the dependence of $N_{\rm d}$ on the amplitude
(top panels) and shape (bottom panels) of the initial power spectrum $P(k)$.
We follow Miralda-Escud\'e et al.'s (1996) suggestion of plotting
$N_{\rm d}$ against $F(D)$, the fraction of the spectrum with
flux decrement less than $D$, which makes the model predictions
nearly independent of the constant $A$ (equation~\ref{eqn:fgpa}) that
relates optical depth to overdensity.
Like the 3-d genus, the threshold crossing frequency is sensitive
to the shape of $P(k)$, with a bluer power spectrum leading to
choppier \lya forest spectra and hence to higher $N_{\rm d}$.
When the fluctuation amplitude is small, 
the threshold crossing frequency is independent of the amplitude of $P(k)$,
but at higher fluctuation amplitudes non-linear merging of structures
tends to depress $N_{\rm d}$, again analogous to the behavior of the 3-d
genus curve.  

Compared to the flux decrement distribution,
the threshold crossing frequency is more sensitive to numerical simulation
parameters, as shown by
the tests on PM simulations in Figure~\ref{fig:dcrossres}.
These tests suggest that our current SPH simulations 
significantly underestimate $N_{\rm d}$
at $25\kms$ because of their finite mass resolution and slightly
overestimate $N_{\rm d}$ at $100\kms$ because of their finite box size.

Figure~\ref{fig:dcrossdata} compares results from SPH simulations to
measurements from the HIRES sample.  The CCDM model predicts the
lowest value of $N_{\rm d}$ because of its high fluctuation amplitude,
while the OCDM and (especially) LCDM models predict higher $N_{\rm d}$
because of the influence of $\om$ and $\ol$ on
the ratio $H(z)/H_0$, which determines the relation between
comoving $\hmpc$ and the observable units of $\kms$ (recall that
$N_{\rm d}$ is measured per unit redshift).  
Because of the numerical uncertainties indicated by Figure~\ref{fig:dcrossres},
we can draw few conclusions from the simulation-data comparison at
present, except to say that the model predictions are in the right ballpark
at $100\kms$.  The different shape of the observed $N_{\rm d}$ curve at
$25\kms$ may well be an effect of noise, which adds spurious threshold
crossings at low flux decrements (the leftmost data point is for $D=0.04$).
A useful comparison at this scale will require careful investigation
of noise and continuum-fitting effects, and perhaps restriction of the
sample to the data with the highest signal-to-noise ratio.

\section{Recovery of the Mass Power Spectrum}

Another obvious continuous field statistic is the flux power spectrum.
As one might expect from the analogy between equation~\eqref{flux}
and ``local'' models of biased galaxy formation 
(Coles 1993; Fry \& Gazta\~naga 1993; Scherrer \& Weinberg 1998),
the shape of the flux power spectrum on large scales is the same
as the shape of the underlying mass power spectrum $P(k)$.
On moderately non-linear scales, it resembles the shape of the
linear mass $P(k)$ rather than the non-linear $P(k)$.
Furthermore, if one chooses the parameter $A$ in order to match
the observed
$\langle D \rangle$, then the amplitude of the flux power spectrum
depends only on the amplitude of the mass power spectrum.
Croft et al.\ (1998a, hereafter CWKH) exploit these facts 
in a procedure that recovers
the linear mass power spectrum directly from \lya forest data.
Once the shape of $P(k)$ is measured from the flux power spectrum
(or, more precisely, from the power spectrum of the ``Gaussianized'' flux;
see CWKH), the amplitude is determined using PM simulations with
Gaussian initial conditions and this $P(k)$ shape, choosing the
amplitude for which the simulations reproduce the observed amplitude
of the flux power spectrum.
In tests on artificial spectra from SPH simulations, the CWKH method
correctly recovers the shape and amplitude of the true, linear mass
power spectrum on scales $\lambda \ga 1\hmpc$ (comoving).

\begin{figure*}
\centerline{
\epsfxsize=3.6truein
\epsfbox[60 470 555 735]{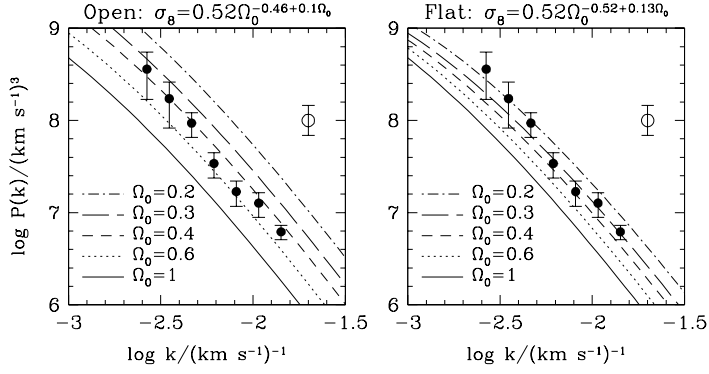}
}
\caption[]{
\label{fig:pkgamma}
Constraints on $\om$ obtained by combining the CWPHK
measurement of $P(k)$ at $z=2.5$ with the mass
function of galaxy clusters at $z=0$, for $\ol=0$ (left)
and $\ol=1-\om$ (right).
Filled circles with $1\sigma$ error bars show the measured
$P(k)$.
The error bar on the open
circle indicates the normalization uncertainty: at the $1\sigma$ level,
all points can be shifted coherently up or down by this amount.
Curves show $P(k)$ at $z=2.5$ for cluster-normalized models with
a power spectrum shape parameter $\Gamma=0.2$ and
various values of $\om$, as indicated.  
The cluster normalization constraints, from 
Eke et al.\ (1996), are listed above each panel.
}
\end{figure*}

We have recently applied this technique to a set of
19 QSO spectra, obtained between 1987 and 1994,
originally for the purpose of studying chemical abundances in 
damped \lya systems (Pettini et al.\ 1994, 1997).
The resolution of these spectra ranges from
$\sim 0.8-2.3$\AA\ (typically $\sim 1.5$\AA), and the signal-to-noise
ratio ranges from $\sim 10-90$ (with typical ${\rm S/N}\geq 40$).  
The analysis is described in detail by Croft et al.\ (1998b, hereafter
CWPHK).

In gravitational instability models with Gaussian initial conditions,
matching the observed mass function of galaxy clusters requires
$\sigma_8 \om^{0.5} \approx 0.5$ (White, Efstathiou, \& Frenk 1993;
Eke, Cole, \& Frenk 1996).  Figure~\ref{fig:pkgamma}, from
Weinberg et al.\ (1998a), compares the CWPHK determination
of $P(k)$ at $z=2.5$ to the predictions of cluster-normalized
models with various values of $\om$.  We adopt the $P(k)$ shape
parameter $\Gamma=0.2$ favored by studies of large scale galaxy clustering.
Models with high $\om$ have low $\sigma_8$ and predict a $P(k)$ that 
is too low to match the \lya forest results.
Models with low $\om$ have high $\sigma_8$ and predict a $P(k)$ that
is too high.
In Weinberg et al.\ (1998a), we formalize this argument to obtain 
constraints with $1\sigma$($2\sigma$) uncertainties
\begin{eqnarray}
\om &=& 0.46~^{+0.12(0.29)}_{-0.10(0.17)} 
\quad {\rm open} , \label{eqn:omega} \\
\om &=& 0.34~^{+0.13(0.32)}_{-0.09(0.16)} 
\quad {\rm flat} ,  \nonumber
\end{eqnarray}
for $\Gamma=0.2$.
The difference between open and flat models mainly reflects
the dependence of the linear growth factor on $\ol$.
The more general best fit result is $\om+0.2\ol=0.46+1.3(\Gamma-0.2)$.
Higher $\Gamma$ yields higher $\om$ because there is less contribution
to $\sigma_8$ from scales beyond that of the CWPHK measurement.

\begin{figure*}
\centerline{
\epsfxsize=3.6truein
\epsfbox[60 470 555 735]{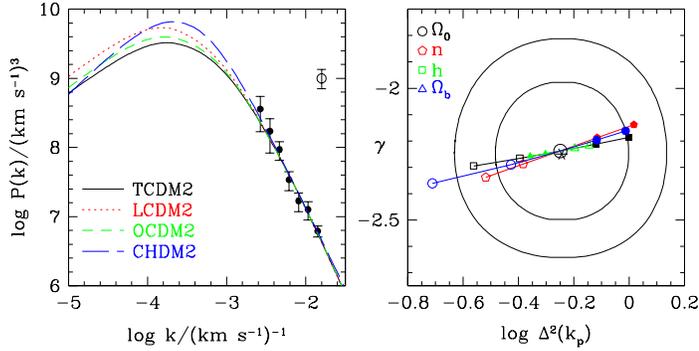}
}
\caption[]{
\label{fig:pkcdm}
Constraints on the parameters of CDM models obtained by combining
the CWPHK $P(k)$ with COBE-DMR normalization.
The left panel shows linear mass
power spectra of four COBE-normalized CDM models
(parameters listed in Table~\ref{tab:2}),
with the inflationary power spectrum index $n$ 
chosen to give the best match to the measured $P(k)$
(filled circles).  The right panel illustrates constraints
on the parameters of the LCDM model.  
The asterisk marks CWPHK's values of the logarithmic 
slope $\gamma$ and amplitude $\Delta^2 \equiv k^3 P(k)/2\pi^2$ of $P(k)$
at wavenumber $k_p=0.008\invkms$. Concentric circles show the 
$1\sigma$ and $2\sigma$ constraints on these parameters. 
Our fiducial version of the LCDM model 
predicts the slope and amplitude indicated by the central open circle.
Filled (open) points show the effect of increasing (decreasing)
one parameter while keeping the others fixed.  Each step corresponds
to $\Delta\om=0.1$ (circles), $\Delta n=0.05$ (pentagons),
$\Delta h=0.05$ (squares), or $\Delta\omb=0.01$ (triangles).
}
\end{figure*}

\begin{table*}
\caption[]{Parameters of the cosmological models shown in 
  Figure~\ref{fig:pkcdm}.  All of these models have 
  $\omb h^2=0.02$.}
  \centering
  \begin{tabular}{l|*{8}{c}}
    \hline
    Model & $\om$ & $\ol$ & $\Omega_\nu$ & $n$ & $h$ & 
	  $\sigma_8$ \\
    \hline\hline
    TCDM2          & 1.0 & 0.0 & 0.0 & 0.84 & 0.50 & 0.92 \\
    LCDM2          & 0.4 & 0.6 & 0.0 & 0.88 & 0.65 & 0.74 \\
    OCDM2          & 0.55 & 0.0 & 0.0 & 0.88 & 0.65 & 0.67 \\
    CHDM2          & 1.0 & 0.0 & 0.2 & 1.10 & 0.50 & 0.96 \\
  \end{tabular}
\label{tab:2}
\end{table*}

The main assumptions behind this $\om$ constraint are gravitational
instability and Gaussian initial conditions.  With the more restrictive
assumptions of inflation and CDM, we can obtain other constraints on
cosmological parameters by combining the \lya $P(k)$ with COBE-DMR
normalization.  Figure~\ref{fig:pkcdm}, based on Phillips et al.\ (1998),
shows power spectra of four COBE-normalized CDM models with the
inflationary power spectrum index $n$ chosen so that
the predicted power spectrum matches the CWPHK result
(see model parameters in Table~\ref{tab:2}).
While all of these models can simultaneously match COBE-DMR and
the \lya forest $P(k)$, the high $\sigma_8$ values of the $\Omega=1$ 
models (TCDM2, CHDM2)
would imply excessively massive galaxy clusters at $z=0$, in agreement 
with our earlier, more general argument.
The right hand panel of Figure~\ref{fig:pkcdm} illustrates the
constraints from COBE and the \lya forest $P(k)$ alone in the case of LCDM.
With our fiducial parameter choices, this model reproduces the slope
and amplitude of the CWPHK $P(k)$ almost perfectly.
Changing $\om$, $n$, $h$, or $\omb$ in isolation would 
change the predicted slope and amplitude as indicated.
Because the different parameter changes
have nearly degenerate effects on the \lya $P(k)$, our measurement
imposes a single constraint on a combination of the parameters:
\begin{eqnarray}
&&\left(\frac{\om}{0.4}\right)  \left(\frac{h}{0.65}\right)^{2.96}
\left(\frac{n}{0.88}\right)^{3.92} \; \times \nonumber \\
&&\left(\frac{\omb h^2}{0.02}\right)^{-0.445} = 1.0 \pm 0.5. 
\label{eqn:cdm}
\end{eqnarray}

As Figure~\ref{fig:pkcdm} shows, parameter choices that reproduce
the measured amplitude of $P(k)$ also reproduce the measured
slope.  This ``coincidence'' represents a significant success of
the inflation+CDM scenario: the CWPHK measurement confirms its
generic prediction of a linear power spectrum that curves steadily
towards $k^{n-4}$ on small scales.

We plan to apply this technique to a sample of 99 moderate resolution
QSO spectra originally obtained by three of us (SB, DK, DT) for the 
purpose of identifying candidate systems for deuterium absorption studies.
The larger sample size will greatly reduce the statistical uncertainties
in $P(k)$, leading to tighter constraints on $\om$ and other parameters.
A preliminary analysis of this sample yields results within the $1\sigma$
errors of the CWPHK measurement.

Figures~\ref{fig:fdfdata}, \ref{fig:dcrossdata}, \ref{fig:pkgamma},
and \ref{fig:pkcdm}, and equations~\eqref{omega} and~\eqref{cdm},
illustrate the power of the \lya forest as a test of cosmological theories.
In contrast to the current situation for galaxies, 
for the \lya forest we can predict the form of the ``bias relation''
(equation~\ref{eqn:flux})
based on simple considerations of photoionization and adiabatic cooling.
The parameter $A$ that determines the strength of this bias
is not well known {\it a priori}, but it
can be measured for any given model using the mean flux decrement, 
leaving all other statistical properties of the forest as independent 
predictions.  While these statements 
are based on the approximate picture of the \lya forest
outlined in \S 2, hydrodynamic simulations can provide the necessary
tests of and corrections to this picture.
The most significant corrections --- peculiar velocities, thermal
broadening, and shock heating --- are straightforward to compute with such
simulations.
The factors that ultimately limit the comparison between models
and data are likely to be observational uncertainties in continuum
fitting, which make it difficult to measure weak fluctuations on large
scales, and theoretical uncertainties in the spatial uniformity
of the $\rho-T$ relation.
At present it appears that these uncertainties 
are small compared to the differences between cosmological models.

What can we expect to learn from the \lya forest over the next few years?
Most clearly, we should get precise constraints on the amplitude and
logarithmic slope of the linear mass power spectrum on
scales $\sim 1000\kms$ at $z\sim 2-3.5$, from application of the CWKH
method to larger data sets and from independent checks with HIRES data
using statistics like the flux decrement distribution and threshold
crossing frequency.  Consistency among these statistical properties will
test the hypothesis of Gaussian initial conditions.  The threshold
crossing frequency provides additional constraints on $\om$ and $\ol$ because
of their influence on the conversion from comoving $\hmpc$ to redshift.
If continuum fitting can be done with sufficient accuracy, it may
be possible to measure curvature of the linear power spectrum or other
departures from a power law shape.
It should be possible to detect the signature of gravitational 
growth of fluctuations over the range $z\sim 2-3.5$.  
Using the rate of growth to constrain $\om$ 
and $\ol$ probably requires accurate $P(k)$ measurements
down to $z\sim 1$, which will be more difficult to achieve because 
such measurements must be based on HST data and because the 
physics of the \lya forest becomes somewhat more complicated at
low redshift (Dav\'e et al.\ 1998).
Measurements of the flux autocorrelation or flux power spectrum
towards QSO pairs can constrain $\ol$ through
its influence on spacetime geometry (CWKH; Hui, Stebbins, \& Burles 1998;
McDonald \& Miralda-Escud\'e 1998).
Finally, all of these results can be combined with complementary
constraints from CMB anisotropies, the cluster mass
function, Type Ia supernovae, and so forth to test the inflation+CDM
scenario of structure formation and to determine the parameters
of the universe that we live in.



\end{document}